\documentstyle[sprocl,epsfig]{article}

\def\PRD{{\em Phys. Rev.} D}
\def\EPJ{{\em Eur. Phys. J.} C }
\def\PRT{\em Phys. Rept.}
\def\APP{{\em Acta Phys. Polon.} B}

\def\be{\begin{equation}}
\def\ee{\end{equation}}
\def\bea{\begin{eqnarray}}
\def\eea{\end{eqnarray}}

\begin{document}

\title{\bf Probing the Majorana Nature and CP Properties of Neutralinos}

\author{S.Y. Choi}

\address{Department of Physics, Chonbuk National University,
         Chonju 561--756, Korea}


\maketitle\abstracts{Two powerful and straightforward methods are
presented for probing the Majorana nature and CP violation of
neutralinos at future $e^+e^-$ linear colliders.}


The search for supersymmetry (SUSY) is one of the main goals at
present and future colliders as SUSY is generally accepted as one
of the most promising concepts for physics beyond the Standard
Model [1]. A characteristic feature of SUSY theories is the
presence of neutralinos $\tilde{\chi}^0$, the spin--1/2 Majorana
superpartners of the neutral gauge and Higgs bosons. The
neutralinos are expected to be among the light supersymmetric
particles that can be produced copiously at future high--energy
colliders. Once several neutralino candidates are observed at such
high--energy collider experiments it will be crucial to establish
the Majorana nature and CP properties of neutralinos as well as to
reconstruct the fundamental SUSY parameters [2].

In this report, we present two powerful and straightforward
methods for probing the Majorana nature and CP violation in the
neutralino system in the framework of the minimal supersymmetric
standard model (MSSM). One method is based on a combined analysis
of the threshold excitations of neutralino pair production in
$e^+e^-$ annihilation and the fermion invariant mass distribution
near the end point of three--body neutralino fermionic decays
[3,4]. The other method is based on the measurement of $Z$--boson
polarization in the two--body decays $\tilde{\chi}^0_i \to
\tilde{\chi}^0_j\, Z$ [5].

The mixing of the neutral U(1) and SU(2) gauginos, $\tilde{B}$ and
$\tilde{W}^3$, and higgsinos, $\tilde{H}^0_{1,2}$, in the MSSM is
described by a $4\times 4$ matrix $N$, diagonalizing the $4\times
4$ symmetric mass matrix. The relevant fundamental SUSY parameters
are the U(1) and SU(2) gaugino mass parameters, $M_1$ and $M_2$,
the higgsino mass parameter $\mu$ and $\tan\beta=v_2/v_1$, the
ratio of the two Higgs vacuum expectation values. In general the
mass parameters $M_1$, $M_2$ and $\mu$ are complex. By
re--parameterizing the fields, $M_2$ can be taken real and
positive, while the parameter $M_1$ is assigned the phase $\Phi_1$
and the parameter $\mu$ the phase $\Phi_\mu$. For the phases
different from $0$ and $\pi$ the neutralino system is CP
noninvariant.

When the electron and fermion masses are neglected both the
production processes, $e^+e^- \to
\tilde{\chi}^0_i\tilde{\chi}^0_j$, near threshold and the
three--body decays, $\tilde{\chi}^0_i \to \tilde{\chi}^0_j f
\bar{f}$ near the fermion invariant mass end point are effectively
regarded as processes of a static vector or axial--vector current
exchange between two neutralinos. In the CP invariant case, the
production of a neutralino $\{ij\}$ pair and the neutralino decay
$\tilde{\chi}^0_i \to \tilde{\chi}^0_j\, V$ through a vector or
axial--vector current with positive intrinsic CP parity satisfy
the CP relations
\begin{eqnarray}
&& 1\, =\, +\eta^i \eta^j\, \left(-1\right)^L
   \label{eq:prod_selection} \\
&& 1\, =\, -\eta^i \eta^j\, \left(-1\right)^L
   \label{eq:decay_selection}
\end{eqnarray}
in the static limit of two neutralinos, where $\eta^i=\pm i$ is
the intrinsic CP parity of $\tilde{\chi}^0_i$ and $L$ is the
orbital angular momentum of the produced neutralino $\{ij\}$ pair
and of the final state of $\tilde{\chi}^0_j$ and $V$,
respectively. The selection rules (\ref{eq:prod_selection},
\ref{eq:decay_selection}) reflect that if two neutralinos
$\tilde{\chi}^0_i$ and $\tilde{\chi}^0_j$ have the same or
opposite CP parity, the current for the neutralino pair production
must be pure axial--vector or pure vector form, respectively. The
intrinsic sign difference of the two selection rules is because
two $u$--spinors are associated with the currents in the
neutralino to neutralino transition, while a $u$ spinor and a $v$
spinor are involved in the neutralino pair production.

One immediate consequence of the selection rules
(\ref{eq:prod_selection},\ref{eq:decay_selection}) is that, in the
CP invariant case, if the production of a pair of neutralinos with
the same (opposite) CP parity through a vector or axial--vector
current is excited slowly in $P$ waves (steeply in $S$ waves),
then the neutralino to neutralino transition via such a vector or
axial--vector current is excited sharply in $S$ waves (slowly in
$P$ waves). In the CP noninvariant case the orbital angular
momentum is, however, no longer restricted by the selection rules.
Consequently, CP violation in the neutralino system can clearly be
signalled in two ways by (a) the sharp $S$--wave excitations of
the production of three non--diagonal $\{ij\}$, $\{ik\}$ and
$\{jk\}$ pairs near threshold [2,3] or by (b) the simultaneous
$S$--wave excitations of the production of any non--diagonal
$\{ij\}$ pair in $e^+e^-$ annihilation near threshold and of the
fermion invariant mass distribution of the neutralino three--body
decays $\tilde{\chi}^0_i \to \tilde{\chi}^0_j f\bar{f}$ near the
kinematical end point [4]. It is in particular a great merit that
even if only the two light neutralinos $\tilde{\chi}^0_{1,2}$ are
accessed kinematically at the initial--phase $e^+e^-$ linear
collider, the combined analysis of the production of the
neutralino $\{12\}$ pair and the decay $\tilde{\chi}^0_2 \to
\tilde{\chi}^0_1 f\bar{f}$ allows us to probe CP violation in the
neutralino system. A clear numerical demonstration of the combined
analysis of the processes,
$e^+e^-\rightarrow\tilde{\chi}^0_1\tilde{\chi}^0_2$ and
$\tilde{\chi}^0_2 \to \tilde{\chi}^0_1 l^+ l^-$, is illustrated in
Fig.1 for the parameter set $\{\tan\beta=10, |M_1|=100\, {\rm
GeV}, M_2=150\, {\rm GeV}, |\mu|=400\, {\rm GeV}, \Phi_\mu=0;
m_{\tilde{l}_{L,R}}=250/200\, {\rm GeV}\}$.

\begin{figure}[!h]
\begin{center}
\vspace*{.2cm} \epsfig{file=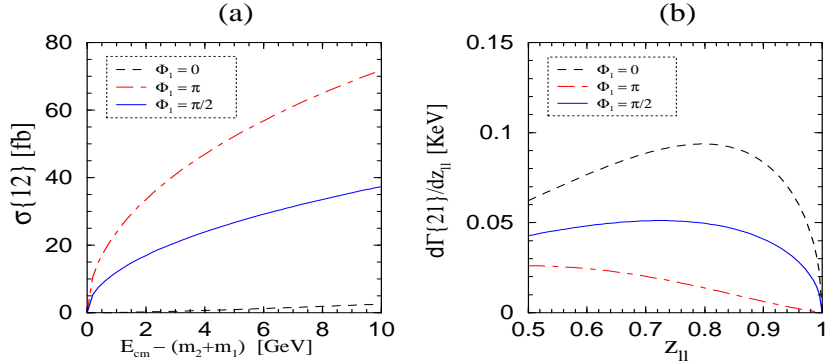,width=11.cm,height=5.cm}
\end{center}
\caption{(a) The threshold behavior of the neutralino production
cross section $\sigma\{12\}$ and (b) the lepton invariant mass
distribution for the parameter set given in the text.}
\label{fig:prod_dec}
\end{figure}

Once two--body decays of a neutralino $\tilde{\chi}^0_i$ into $Z$,
Higgs bosons or sfermions are kinematically open, the combined
production--decay analysis cannot be exploited for probing CP
violation in the neutralino system. Nevertheless, if the two--body
decays $\tilde{\chi}^0_i \to \tilde{\chi}^0_j Z$ is not too
strongly suppressed, the $Z$ polarization reconstructed via
leptonic $Z$--boson decays with great precision allows us to probe
the Majorana nature and CP violation in the neutralino system [5].
The Majorana nature of neutralinos forces the vector and axial
vector $Z\tilde{\chi}^0\tilde{\chi}^0$ couplings to be pure
imaginary and pure real, respectively. This characteristic
Majorana property leads to one important relation between the
decay widths with the $Z$--boson helicities $\pm 1$:
$\Gamma[\tilde{\chi}^0_i \to \tilde{\chi}^0_j Z(+)]=
\Gamma[\tilde{\chi}^0_i \to \tilde{\chi}^0_j Z(-)]$, that is valid
even in the CP noninvariant case. In addition, CP violation in the
neutralino system can be probed by measuring the ratio of the
longitudinal to transverse decay widths, if the relevant
neutralino masses are measured with good precision, independently
of the decay modes.

The suggested methods  can be exploited experimentally in a
straightforward manner so that they are expected to provide us
with first--stage indications of the Majorana nature and CP
violation in the neutralino system.

\section*{References}

\end{document}